\documentclass[12pt]{article}
\usepackage[english]{babel}
\usepackage{amsmath}
\usepackage{color}
\usepackage[titletoc,toc,title]{appendix}
\Roman{section}
\newcommand{\eq}[1]{\begin{equation}#1\end{equation}}
\newcommand{\naw}[1]{\left(#1\right)}
\newcommand{\ket}[1]{\left|#1\right>}
\newcommand{\bra}[1]{\left<#1\right|}
\newcommand{\av}[1]{\left<#1\right>}
\newcommand{\com}[1]{\left[#1\right]}
\newcommand{\modu}[1]{\left|#1\right|}

\begin{document}

\begin{center}
\textsc{\Large{ENTANGLERS IN 3-STRATEGIES ELW GAMES}}

\emph{Katarzyna Bolonek-Laso\'n\footnote{kbolonek1@wp.pl}\\ Faculty of Economics and Sociology, Department of Statistical Methods\\ University of Lodz, Poland.}

\end{center}
\begin{abstract}
We construct the general gate operator for 2-players 3-strategies ELW games. It is shown that such a gate, implementing classical strategies, can be constructed out of the elements of Cartan subalgebra of $SU(3)$. The relation between the degree of entanglement and the structure of stability subgroup of initial state is analyzed.
\end{abstract}

\section{Introduction}
In two important papers \cite{EisertWL}, \cite{EisertW} Eisert, Wilkens and Lewenstein proposed the method which allows, given some classical game, to construct its quantum counterpart. The example they described provides a paradigm of quantum game. Since then the theory of quantum games has been a subject of intensive research \cite{Meyer}$\div$\cite{Avishay}.

In their attempt to justify the interest in quantum games Eisert, Lewenstein and Wilkens speculate that games of survival are being played already on molecular level where things are happening according to the rules of quantum mechanics. They also pointed out that there is an intimate connection between the theory of games and the theory of quantum communication.

Any quantum game can be played classically being modelled on a classical computer. However, it can happen that this is not physically feasible due to limited resources and time; in such a case only quantum mechanics allows for an implementation of the game due to the existence of specifical quantum correlations which break the Bell-like inequalities (much like in the domain of quantum computing). 

In the original ELW proposal the set of allowed strategies of each player does not form a group. It has been argued \cite{BenjaminHay} that it is more natural to assume the set of strategies to be closed under multiplication. We adopt here this point of view. 

The original ELW game results from quantization of two-players two-strategies classical game. However, one can consider arbitrary $N$-strategies game as a starting point. It appears then that for constructing reasonable quantum extensions. The key element in the construction of quantum game is the gate operator which provides an entanglement of the initial state. When $N$ grows the number of arbitrary parameters entering the definition of gate operator also grows \cite{Bolonek1} leading to new phenomena.

In the present paper we consider the construction of gate operator in the $N=3$ case. We assume that the set of admissible strategies of each player is the whole $SU(3)$ group. Three parameter family of gate operators is constructed. The cases corresponding to various degrees of entanglement are identified and related to the structure of the stability group of initial state.

The paper is organized as follows. In Sec. II some general considerations are presented. In particular we discuss the structure of stability subgroups of the initial state; we show that the maximal entanglement makes the game essentially trivial \cite{Bolonek2}. Sec. III is devoted to the explicit construction of the gate operator. In the number of cases we compute the generators of stability group confirming the general arguments of Sec. II. Sec. IV is devoted to brief discussion.

Let us note that the three-strategies case differs from original ELW game also by the fact that we can choose a nontrivial subgroup of $SU(3)$, namely $SO(3)$, as the set of admissible strategies. This game will be considered elsewhere \cite{Bolonek3}.       

\section{General considerations}
We start with classical two-players three-strategies game defined by a $3\times 3$ payoff matrix. ELW quantization is performed as follows. To any player (Alice and Bob) the three dimensional Hilbert space ia ascribed which is spanned by the vectors
\eq{\ket{1}=\left (\begin{array}{c} 1\\0\\0 \end{array}\right),\qquad \ket{2}=\left (\begin{array}{c} 0\\1\\0 \end{array}\right),\qquad \ket{3}=\left (\begin{array}{c} 0\\0\\1 \end{array}\right).}

One begins with the vector $\ket{1}\otimes\ket{1}$. The entanglement is provided by a reversible gate operator $\hat{J}$ which plays the main role in quantization procedure. The initial state of the game is
\eq{\ket{\Psi_i}=\hat{J}\naw{\ket{1}\otimes\ket{1}}.\label{f}}
The set of strategies at the disposal of each player is a subset of $SU(3)$ manifold. In fact, in the main body of the paper we assume that it is a whole $SU(3)$ group. The choice of the admissible strategies is an important element of the definition of quantum game which determines some of its main properties like, for example, the existence of Nash equilibria. It seems natural to assume the set of strategies to be closed under matrix multiplication (although this was not the case in the original ELW paper); admissible strategies form a group. 
The whole $SU(3)$ group is a simplest choice. However, there is another possibility: three dimensional representation of $SU(2)$ group. Assuming further irreducibility we are dealing with $SO(3)$ embedding into $SU(3)$. 

The above reasoning is applicable also to general $N$-strategies game. Any compact Lie group admitting N-dimensional unitary representation can serve as a strategy manifold for individual player. \\
After the players have performed their moves $U_A$, $U_B$, the final measurement is made yielding the final state of the game
\eq{\ket{\Psi_f}=\hat{J}^+\naw{U_A\otimes U_B}\hat{J}\naw{\ket{1}\otimes\ket{1}}.}
This allows us to compute the players expected payoffs
\eq{\$^{A,B}=\sum_{\sigma,\sigma'=1}^{3}p_{\sigma\sigma'}^{A,B}\modu{\av{\sigma,\sigma'|\Psi_f}}^2.}

The gate operator $\hat{J}$ introduces entanglement into the initial state (\ref{f}) providing the game genuinely quantum character. In order to study the influence of the entanglement we put
\eq{\Psi_i=F_{ij}\ket{i}\otimes\ket{j}\label{f5}} 
where $F_{ij}$ is symmetric (we are considering symmetric game) and the summation over repeated indices is understood. The entanglement of $\ket{\Psi_i}$ can be studied by taking the partial trace (with respect to arbitrary player) of the initial density operator 
\eq{\rho_i=\ket{\Psi_i}\bra{\Psi_i}.}
One obtains 
\eq{\text{Tr}_B\rho_i=\naw{FF^+}_{ij}\ket{i}\bra{j}.}
The maximal entanglement of the initial state corresponds to
\eq{FF^+=\frac{1}{3}I}
i.e. $\widetilde{F}\equiv\sqrt{3}F$ is unitary.

Let us now determine the stability subgroup of $\ket{\Psi_i}$ in $SU(3)\times SU(3)$ in the case of maximal entanglement. By applying $U_A\otimes U_B$ to $\ket{\Psi_i}$ one finds
\eq{\naw{U_A\otimes U_B}\ket{\Psi_i}=\naw{U_AFU_B^T}_{kl}\ket{k}\otimes\ket{l}.}
The invariance condition reads
\eq{\widetilde{F}=U_A\widetilde{F}U_B^T}
with the general solution
\eq{\begin{split}
&U_A=U\\
&U_B=\widetilde{F}\overline{U}\widetilde{F}^+
\end{split}\label{f3}}
$U\in SU(3)$ being an arbitrary matrix. So the stability subgroup is, up to an automorphism, the diagonal subgroup of $SU(3)\times SU(3)$. This means that the manifold of strategies of $\underline{\text{both}}$ players is isomorphic to $SU(3)\times SU(3)/diag\naw{SU(3)\times SU(3)}$ i.e. it is eight dimensional.

The maximal entanglement implies that to any strategy of the first player there exist an appropriate counterstrategy of the second one. In fact, let $\naw{U_A,U_B}$ be a pair of arbitrary strategies; one can write the decomposition
\eq{\naw{U_A,U_B}=\naw{V,U_B\widetilde{F}U_A^T\overline{V}\widetilde{F}^+}\naw{V^+U_A,\widetilde{F}V^T\overline{U}_A\widetilde{F}^+}\label{a}}
V being arbitrary element of SU(3). 

Eq. (\ref{a}) has the following interpretation. The second factor on the right hand side belongs to the stability subgroup. Let V be the Alice actual move; Bob wants to obtain the payoff resulting from the pair of moves $\naw{U_A,U_B}$. Then $U_B\widetilde{F}U_A^T\overline{V}\widetilde{F}^+$ is its appropriate countermove.

The degree of entanglement and the structure of stability subgroup depend on $F$. The invariance condition
\eq{U_AFU_B^T=F\label{ab}}
implies
\eq{U_A\naw{FF^+}U_A^+=FF^+.}
The structura of $U_A$ depends on the eigenvalues of $FF^+$. Three equal ones define the maximal entanglement described above. The other possibilities are:
\begin{description}
\item{(i)} two equal eigenvalues: then $U_A$ belongs to $S\naw{U(2)\times U(1)}$ subgroup of $SU(3)$ which is four dimensional; the manifold of strategies for $\underline{\text{both}}$ players is twelve dimensional
\item{(ii)} three different eigenvalues: $U_A$ belongs to $S\naw{U(1)\times U(1)\times U(1)}$ subgroup of $SU(3)$ which is two dimensional; the manifold of strategies for $\underline{\text{both}}$ players is fourteen dimensional. 
\end{description} 
To see that the assumptions concerning the number of eigenvalues of $FF^+$ imply the appropriate structure of the stability subgroup we invoke the polar decomposition theorem which imply the following decomposition of $F$
\eq{F=UDV\label{f1}}
where $U,V\in SU(3)$ and $D$ is hermitean and diagonal. In the case (i) $D$ has two equal eigenvalues; therefore, the set of unitary matrices $W\in SU(3)$ obeying 
\eq{WDW^+=D\label{f2}}
form $S\naw{U(2)\times U(1)}$ group. Moreover, using (\ref{f1}) and (\ref{f2}) we find that general $U_A$, $U_B$ obeying (\ref{ab}) have the form 
\eq{\begin{split}
& U_A=UWU^+\\
&  U_B=V^T\overline{W}\,\overline{V}.
\end{split}}
Similar reasoning applies to the case (iii).      

\section{The gate operator}
Consider two-players three-strategies symmetric game. The gate operator $\hat{J}$ is an unitary operator acting in $H\otimes H$ and obeying $\varsigma\naw{\hat{J}}=\hat{J}$ where $\varsigma$ is the transposition operator, $\varsigma\naw{\varphi\otimes\psi}=\psi\otimes\varphi$. We assume that the classical strategies are implemented in quantum game. To this end we demand that there exist unitary matrices $U_k$; $k=1,2,3$ such that 
\eq{\begin{split}
& U_k\ket{1}=e^{i\varphi_k}\ket{k}\\
& \com{\hat{J},U_j\otimes U_k}=0, \qquad j,k=1,2,3.
\end{split}\label{aa}}
In order to leave as much freedom as possible for the choice of $\hat{J}$ we demand further
\eq{U_iU_j=U_jU_i,\qquad i,j=1,2,3.\label{b}}
For simplicity we take $U_1=I$. First eq. (\ref{aa}) give the following general form of $U_2$ and $U_3$
\eq{U_2=\left (\begin{array}{ccc}
0 & \alpha & \beta\\
e^{i\varphi_2} & 0 & 0\\
0 & \overline{\beta}e^{-i\varphi_2} & -\overline{\alpha}e^{-i\varphi_2}\end{array}\right), \qquad \modu{\alpha}^2+\modu{\beta}^2=1\label{b1}}
\eq{U_3=\left (\begin{array}{ccc}
0 & \gamma & \delta\\
0 & -\overline{\delta}e^{-i\varphi_3} & \overline{\gamma}e^{-i\varphi_3}\\
e^{i\varphi_3} & 0 & 0
\end{array}\right), \qquad \modu{\gamma}^2+\modu{\delta}^2=1.\label{b2}}
Inserting (\ref{b1}) and (\ref{b2}) into (\ref{b}) one finds
\eq{U_2=\left(\begin{array}{ccc}
0 & 0 & \varepsilon e^{-i\varphi_3}\\
e^{i\varphi_2} & 0 & 0\\
0 & \overline{\varepsilon}e^{i\naw{\varphi_3-\varphi_2}} & 0 \end{array}\right )}
\eq{U_3=\left(\begin{array}{ccc}
0 & \varepsilon e^{-i\varphi_2} & 0\\
0 & 0 & \overline{\varepsilon}e^{i\naw{\varphi_2-\varphi_3}}\\
e^{i\varphi_3} & 0 & 0 \end{array}\right)}
where $\varepsilon$ is any cubic root from unity. \\
The matrices $U_1$, $U_2$, $U_3$ commute so they have common eigenvectors
\eq{\widetilde{\ket{1}}=\frac{1}{\sqrt{3}}\left(\begin{array}{c} 1 \\ e^{i\varphi_2} \\ \overline{\varepsilon}e^{i\varphi_3}\end{array}\right),\qquad \widetilde{\ket{2}}=\frac{1}{\sqrt{3}}\left(\begin{array}{c} 1 \\ \overline{\varepsilon}e^{i\varphi_2} \\ e^{i\varphi_3}\end{array}\right),\qquad \widetilde{\ket{3}}=\frac{1}{\sqrt{3}}\left(\begin{array}{c} 1 \\ \varepsilon e^{i\varphi_2} \\ \varepsilon e^{i\varphi_3}\end{array}\right).}
They are linearly independent provided $\varepsilon\neq 1$. The corresponding eigenvalues are given in Table 1.
\begin{center}\begin{tabular}{|c|c|c|c|}
 \cline{1-4}
  &$ U_1$ & $U_2$ & $U_3$\\  \cline{1-4}
$\lambda_1$  & 1 & 1 & $\varepsilon$ \\ \cline{1-4}
$\lambda_2$ & 1 & $\varepsilon$ & 1 \\ \cline{1-4}
$\lambda_3$ & 1 & $\varepsilon^2$ & $\varepsilon^2$\\ 
 \hline
\end{tabular}\\
\end{center}
Therefore, denoting
\eq{V=\frac{1}{\sqrt{3}}\left(\begin{array}{ccc}
1 & 1 & 1\\
e^{i\varphi_2} & \overline{\varepsilon} e^{i\varphi_2} & \varepsilon e^{i\varphi_2}\\
\overline{\varepsilon}e^{i\varphi_3} & e^{i\varphi_3} & \varepsilon e^{i\varphi_3}\end{array}\right),\qquad VV^+=I}
one obtains
\eq{\begin{split}
& \widetilde{U}_1=V^+U_1V=I\\
& \widetilde{U}_2=V^+U_2V=diag\naw{1,\varepsilon, \varepsilon^2}\\
& \widetilde{U}_3=V^+U_3V=diag\naw{\varepsilon,1,\varepsilon^2}
\end{split}.\label{f4}}

In order to find a general form of the gate operator $\hat{J}$ we use second eq. (\ref{aa}).\\
On defining
\eq{\widetilde{J}\equiv\naw{V^+\otimes V^+}\hat{J}\naw{V\otimes V}} 
we find that $\widetilde{J}$ commutes with all $\widetilde{U}_j\otimes\widetilde{U}_k$. As a result $\widetilde{J}$ is diagonal and one can take
\eq{\widetilde{J}=\exp i\naw{\tau\naw{\Lambda\otimes\Lambda}+\rho\naw{\Lambda\otimes\Delta+\Delta\otimes\Lambda}+\sigma\naw{\Delta\otimes\Delta}}\label{bb}}
where $\tau$, $\rho$ and $\sigma$ are arbitrary real numbers and
\eq{
\Lambda\equiv\left(\begin{array}{ccc}
1 & 0 & 0\\
0 & -1 & 0\\
0 & 0 & 0\end{array}\right), \qquad \Delta\equiv\left(\begin{array}{ccc}
1 & 0 & 0\\
0 & 0 & 0\\
0 & 0 & -1 \end{array}\right)} 
and we have assumed, without loosing generality, $\det\hat{J}=1$. This is not the most general form as we have neglected terms of the form $I\otimes\Lambda+\Lambda\otimes I$ etc. because they can be accomodated by appropriate redefinition of Alice and Bob strategies. 

Note that the exponent on the right hand side of eq. (\ref{bb}) is a linear combination of symmetrized tensor products of the elements of Cartan subalgebra of $SU(3)$. In fact, denoting by $\lambda_i$, $i=1,...,8$, the standard Gell-Mann matrices we have
\eq{\Lambda=\lambda_3,\qquad \Delta=\frac{1}{2}\naw{\lambda_3+\sqrt{3}\lambda_8}.}
The outcome probabilities
\eq{P_{\sigma\sigma'}\equiv\modu{\bra{\sigma,\sigma'}\hat{J}^+\naw{U_A\otimes U_B}\hat{J}\ket{1,1}}^2}
can be rewritten as
\eq{P_{\sigma\sigma'}=\modu{\bra{\widetilde{\sigma},\widetilde{\sigma}'}\widetilde{J}^+\naw{\widetilde{U}_A\otimes\widetilde{U}_B}\widetilde{J}\ket{\widetilde{1},\widetilde{1}}}^2}
with
\eq{\ket{\widetilde{\sigma},\widetilde{\sigma}'}\equiv\naw{V^+\otimes V^+}\ket{\sigma,\sigma'}.}  
So we can use $\ket{\widetilde{\sigma},\widetilde{\sigma}'}$ vectors (in particular, the game starts $\ket{\widetilde{1},\widetilde{1}}$) which only amounts to relabelling of Alice and Bob strategies. In this context eq. (\ref{bb}) defines the general gate operator.

As it has been mentioned in the previous section that an important role in the ELW game is played by the stability subgroup of the initial state. Its structure is, in turn, determined by the degree of entanglement of the latter. The reduced density matrix reads:
\begin{equation}
\text{Tr}_B\rho_i =\frac{1}{9}\left(\begin{array}{c|c|c}
 &  e^{i\naw{3\rho+\sigma+2\tau}}+ & e^{i\naw{3\rho+2\sigma+\tau}}+\\
 3 & +e^{-i\naw{\rho +2\tau}}+ & +e^{-i\naw{2\rho+\tau}}\\
 & +e^{-i\naw{2\rho+\sigma}} & +e^{-i\naw{\rho+2\sigma}}\\ \hline
  e^{-i\naw{3\rho+\sigma+\tau}}+ &  & e^{i\naw{\sigma-\tau}}+\\
  +e^{i\naw{\rho+2\tau}}+ & 3 & +e^{-i\naw{\rho-\tau}}+\\
  +e^{2\rho+\sigma} & & +e^{i\naw{\rho-\sigma}}\\ \hline
  e^{-i\naw{3\rho+2\sigma+\tau}}+ & e^{-i\naw{\sigma-\tau}} + & \\
  + e^{i\naw{2\rho+\tau}}+ & +e^{i\naw{\rho-\tau}}+ & 3\\
  +e^{i\naw{\rho+2\sigma}} & +e^{-i\naw{\rho-\sigma}} & \\ 
  \end{array}\right) \label{d3}
 \end{equation}

It is shown in Appendix B that the maximal entanglement corresponds to the following sets of the values of parameters $\rho$, $\sigma$, and $\tau$:

\begin{equation}\begin{split}
& \left\{ \begin{array}{c}
\tau=\rho=\sigma-\frac{2\pi}{3}\\
\sigma=\frac{2\pi}{3},\frac{8\pi}{9},\frac{10\pi}{9},\frac{4\pi}{3},\frac{14\pi}{9},\frac{16\pi}{9},2\pi \end{array}\right.\\
& \left\lbrace \begin{array}{c}
\tau=\rho=\sigma+\frac{2\pi}{3}\\
\sigma=0,\frac{2\pi}{9},\frac{4\pi}{9},\frac{2\pi}{3},\frac{8\pi}{9},\frac{10\pi}{9},\frac{4\pi}{3},\frac{14\pi}{9},\frac{16\pi}{9} \end{array}\right.\\
& \left\{ \begin{array}{c}
\tau=\sigma-\frac{2\pi}{3}\\
\rho=\sigma=\frac{2\pi}{3},\frac{8\pi}{9},\frac{10\pi}{9},\frac{4\pi}{3},\frac{14\pi}{9},\frac{16\pi}{9},2\pi\end{array}\right. \\
&\left\lbrace \begin{array}{c}
\tau=\sigma+\frac{2\pi}{3}\\
\rho=\sigma=0,\frac{2\pi}{9},\frac{4\pi}{9},\frac{2\pi}{3},\frac{8\pi}{9},\frac{10\pi}{9},\frac{4\pi}{3},\frac{14\pi}{9},\frac{16\pi}{9} \end{array}\right.\\
& \left\{ \begin{array}{c}
\rho=\sigma-\frac{2\pi}{3}\\
\tau=\sigma=\frac{2\pi}{3},\frac{8\pi}{9},\frac{10\pi}{9},\frac{4\pi}{3},\frac{14\pi}{9},\frac{16\pi}{9},2\pi \end{array}\right.\\
& \left\lbrace \begin{array}{c}
\rho=\sigma+\frac{2\pi}{3}\\
\tau=\sigma=0,\frac{2\pi}{9},\frac{4\pi}{9},\frac{2\pi}{3},\frac{8\pi}{9},\frac{10\pi}{9},\frac{4\pi}{3},\frac{14\pi}{9},\frac{16\pi}{9} \end{array}\right.
\end{split}\label{d1}
\end{equation}
The generators of the stability subgroup, isomorphic, in the case of maximal entanglement, to $SU(3)$ group, can be obtained using the general considerations of Sec. II. Namely, according to the eq. (\ref{f3}) any generator can be written in the form 
\begin{equation}
X\otimes I+I\otimes\widetilde{F}\overline{X}\widetilde{F}^+\label{d}
\end{equation}  
where $\widetilde{F}$ is the unitary matrix defined below eq. (\ref{f5}); all generators can be obtained by taking as $X$ all Gell-Mann matrices (conventionally divided by two). The explicit form of the operators (\ref{d}) depends on $\widetilde{F}$, i.e. the form of gate operator $\hat{J}$.\\
Alternatively, to find the explicit form of the generators of stability subgroup one can use the direct method described in Appendix A. As it is explained there all the generators can be written in form
\begin{equation}
X\otimes I\pm I\otimes X
\end{equation}
with $X$ appropriately chosen. Following the method of Appendix A we have computed the generators for some of the solutions listed in eq. (\ref{d1}).
\begin{footnotesize}
\begin{description}
\item{(i)} $\rho=\frac{2\pi}{3}$, $\sigma=\tau=0$
\begin{equation}
\begin{split}
& G_1=\naw{\lambda_1-\sqrt{3}\lambda_2+\frac{2}{\sqrt{3}}\lambda_8}\otimes I-I\otimes \naw{\lambda_1-\sqrt{3}\lambda_2+\frac{2}{\sqrt{3}}\lambda_8}\\
& G_2=\naw{\sqrt{3}\lambda_2+\lambda_3+\lambda_4-\frac{1}{\sqrt{3}}\lambda_8}\otimes I-I\otimes \naw{\sqrt{3}\lambda_2+\lambda_3+\lambda_4-\frac{1}{\sqrt{3}}\lambda_8}\\
& G_3=\naw{\lambda_3+2\lambda_6+\frac{1}{\sqrt{3}}\lambda_8}\otimes I-I\otimes\naw{\lambda_3+2\lambda_6+\frac{1}{\sqrt{3}}\lambda_8}\\
& G_4=\naw{\lambda_2+\lambda_5}\otimes I-I\otimes \naw{\lambda_2+\lambda_5}\\
& G_5=\naw{4\lambda_2+\sqrt{3}\lambda_3+2\lambda_7-3\lambda_8}\otimes I-I\otimes\naw{4\lambda_2+\sqrt{3}\lambda_3+2\lambda_7-3\lambda_8}\\
& G_6=\naw{\lambda_1-\frac{1}{2}\lambda_4+\frac{1}{4}\lambda_6-\frac{3\sqrt{3}}{4}\lambda_7-\frac{\sqrt{3}}{2}\lambda_8}\otimes I+\\
&\qquad+I\otimes\naw{\lambda_1-\frac{1}{2}\lambda_4+\frac{1}{4}\lambda_6-\frac{3\sqrt{3}}{4}\lambda_7-\frac{\sqrt{3}}{2}\lambda_8}\\
& G_7=\naw{\lambda_2-\frac{\sqrt{3}}{2}\lambda_4-\lambda_5-\frac{\sqrt{3}}{4}\lambda_6+\frac{1}{4}\lambda_7+\frac{3}{2}\lambda_8}\otimes I+\\
&\qquad +I\otimes\naw{\lambda_2-\frac{\sqrt{3}}{2}\lambda_4-\lambda_5-\frac{\sqrt{3}}{4}\lambda_6+\frac{1}{4}\lambda_7+\frac{3}{2}\lambda_8}\\
& G_8=\naw{\lambda_3-\lambda_4-\frac{1}{2}\lambda_6-\frac{\sqrt{3}}{2}\lambda_7}\otimes I+I\otimes \naw{\lambda_3-\lambda_4-\frac{1}{2}\lambda_6-\frac{\sqrt{3}}{2}\lambda_7}
\end{split}
\end{equation}
\item{(ii)} $\sigma=\frac{2\pi}{3}$, $\rho=\tau=0$
\begin{equation}
\begin{split}
& G_1=\naw{\lambda_1-\sqrt{3}\lambda_7+\frac{2}{\sqrt{3}}\lambda_8}\otimes I-I\otimes\naw{\lambda_1-\sqrt{3}\lambda_7+\frac{2}{\sqrt{3}}\lambda_8}\\
& G_2=\naw{-\lambda_3+2\lambda_4+\frac{1}{\sqrt{3}}\lambda_8}\otimes I-I\otimes\naw{-\lambda_3+2\lambda_4+\frac{1}{\sqrt{3}}\lambda_8}\\
& G_3=\naw{-\lambda_3+\lambda_6+\sqrt{3}\lambda_7-\frac{1}{\sqrt{3}}\lambda_8}\otimes I-I\otimes\naw{-\lambda_3+\lambda_6+\sqrt{3}\lambda_7-\frac{1}{\sqrt{3}}\lambda_8}\\
& G_4=\naw{\lambda_2-\lambda_7}\otimes I-I\otimes\naw{\lambda_2-\lambda_7}\\
& G_5=\naw{\sqrt{3}\lambda_3+2\lambda_5-4\lambda_7+3\lambda_8}\otimes I-I\otimes\naw{\sqrt{3}\lambda_3+2\lambda_5-4\lambda_7+3\lambda_8}\\
& G_6=\naw{\lambda_1+\frac{1}{4}\lambda_4+\frac{3\sqrt{3}}{4}\lambda_5-\frac{1}{2}\lambda_6-\frac{\sqrt{3}}{2}\lambda_8}\otimes I+\\
& \qquad +I\otimes\naw{\lambda_1+\frac{1}{4}\lambda_4+\frac{3\sqrt{3}}{4}\lambda_5-\frac{1}{2}\lambda_6-\frac{\sqrt{3}}{2}\lambda_8}\\
& G_7=\naw{\lambda_2-\frac{\sqrt{3}}{4}\lambda_4-\frac{1}{4}\lambda_5-\frac{\sqrt{3}}{2}\lambda_6+\lambda_7+\frac{3}{2}\lambda_8}\otimes I+\\
&\qquad +I\otimes\naw{\lambda_2-\frac{\sqrt{3}}{4}\lambda_4-\frac{1}{4}\lambda_5-\frac{\sqrt{3}}{2}\lambda_6+\lambda_7+\frac{3}{2}\lambda_8}\\
&G_8=\naw{\lambda_3+\frac{1}{2}\lambda_4-\frac{\sqrt{3}}{2}\lambda_5+\lambda_6}\otimes I+I\otimes\naw{\lambda_3+\frac{1}{2}\lambda_4-\frac{\sqrt{3}}{2}\lambda_5+\lambda_6}
\end{split}
\end{equation}
\item{(iii)} $\tau=\frac{2\pi}{3}$, $\rho=\sigma=0$
\begin{equation}
\begin{split}
& G_1=\naw{\lambda_1-\frac{1}{\sqrt{3}}\lambda_8}\otimes I-I\otimes\naw{\lambda_1-\frac{1}{\sqrt{3}}\lambda_8}\\
& G_2=\naw{\lambda_5+\lambda_7}\otimes I-I\otimes\naw{\lambda_5+\lambda_7}\\
& G_3=\naw{\lambda_2+\sqrt{3}\lambda_3-2\lambda_5}\otimes I-I\otimes\naw{\lambda_2+\sqrt{3}\lambda_3-2\lambda_5}\\
& G_4=\naw{\lambda_4+\lambda_6-\frac{2}{\sqrt{3}}\lambda_8}\otimes I-I\otimes\naw{\lambda_4+\lambda_6-\frac{2}{\sqrt{3}}\lambda_8}\\
& G_5=\naw{2\lambda_3+\lambda_4-2\sqrt{3}\lambda_5-\lambda_6}\otimes I-I\otimes\naw{2\lambda_3+\lambda_4-2\sqrt{3}\lambda_5-\lambda_6}\\
& G_6=\naw{\lambda_1+\lambda_4+\lambda_6+\sqrt{3}\lambda_8}\otimes I+I\otimes\naw{\lambda_1+\lambda_4+\lambda_6+\sqrt{3}\lambda_8}\\
& G_7=\naw{\lambda_2+\frac{\sqrt{3}}{2}\lambda_4+\frac{1}{2}\lambda_5-\frac{\sqrt{3}}{2}\lambda_6-\frac{1}{2}\lambda_7}\otimes I+\\
& \qquad +I\otimes \naw{\lambda_2+\frac{\sqrt{3}}{2}\lambda_4+\frac{1}{2}\lambda_5-\frac{\sqrt{3}}{2}\lambda_6-\frac{1}{2}\lambda_7}\\
& G_8=\naw{\lambda_3+\frac{1}{2}\lambda_4+\frac{\sqrt{3}}{2}\lambda_5-\frac{1}{2}\lambda_6-\frac{\sqrt{3}}{2}\lambda_7}\otimes I+\\
&\qquad +I\otimes\naw{\lambda_3+\frac{1}{2}\lambda_4+\frac{\sqrt{3}}{2}\lambda_5-\frac{1}{2}\lambda_6-\frac{\sqrt{3}}{2}\lambda_7}
\end{split}
\end{equation}
\item{(iv)} $\rho=\frac{4\pi}{3}$, $\sigma=\tau=\frac{2\pi}{3}$
\begin{equation}
\begin{split}
& G_1=\naw{\lambda_1+\sqrt{3}\lambda_2+\frac{2}{\sqrt{3}}\lambda_8}\otimes I-I\otimes\naw{\lambda_1+\sqrt{3}\lambda_2+\frac{2}{\sqrt{3}}\lambda_8}\\
& G_2= \naw{-\sqrt{3}\lambda_2+\lambda_3+\lambda_4-\frac{1}{\sqrt{3}}\lambda_8}\otimes I-I\otimes\naw{-\sqrt{3}\lambda_2+\lambda_3+\lambda_4-\frac{1}{\sqrt{3}}\lambda_8}\\
& G_3=\naw{\lambda_3+2\lambda_6+\frac{1}{\sqrt{3}}\lambda_8}\otimes I-I\otimes\naw{\lambda_3+2\lambda_6+\frac{1}{\sqrt{3}}\lambda_8}\\
& G_4=\naw{\lambda_2+\lambda_5}\otimes I-I\otimes\naw{\lambda_2+\lambda_5}\\
& G_5=\naw{4\lambda_2-\sqrt{3}\lambda_3+2\lambda_7+3\lambda_8}\otimes I-I\otimes\naw{4\lambda_2-\sqrt{3}\lambda_3+2\lambda_7+3\lambda_8}\\
& G_6=\naw{\lambda_1-\frac{1}{2}\lambda_4+\frac{1}{4}\lambda_6+\frac{3\sqrt{3}}{4}\lambda_7-\frac{\sqrt{3}}{2}\lambda_8}\otimes I+\\
& \qquad +I\otimes\naw{\lambda_1-\frac{1}{2}\lambda_4+\frac{1}{4}\lambda_6+\frac{3\sqrt{3}}{4}\lambda_7-\frac{\sqrt{3}}{2}\lambda_8}\\
& G_7=\naw{\lambda_2+\frac{\sqrt{3}}{2}\lambda_4-\lambda_5+\frac{\sqrt{3}}{4}\lambda_6+\frac{1}{4}\lambda_7-\frac{3}{2}\lambda_8}\otimes I+\\
& \qquad +I\otimes\naw{\lambda_2+\frac{\sqrt{3}}{2}\lambda_4-\lambda_5+\frac{\sqrt{3}}{4}\lambda_6+\frac{1}{4}\lambda_7-\frac{3}{2}\lambda_8}\\
& G_8=\naw{\lambda_3-\lambda_4-\frac{1}{2}\lambda_6+\frac{\sqrt{3}}{2}\lambda_7}\otimes I+I\otimes \naw{\lambda_3-\lambda_4-\frac{1}{2}\lambda_6+\frac{\sqrt{3}}{2}\lambda_7}.
\end{split}
\end{equation}
\end{description}
\end{footnotesize}
In all cases one can check that the generators are independent and have the form (\ref{d}) with appropriate $\widetilde{F}$.\\
Next, consider the case when two eigenvalues of the reduced density matrix (\ref{d3}) are equal. The necessary and sufficient conditions for this to be the case are given in Appendix B. When expressed in terms of the initial parameters $\rho$, $\sigma$ and $\tau$ they become quite complicated. Therefore, we shall consider only the cases when only one of them is nonvanishing. Under such a condition the full set of solutions read
\begin{equation}
 \left\{\begin{array}{c}
\sigma=\tau=0\\
\rho=\frac{\pi}{3},\pi,\frac{5\pi}{3}
\end{array}\right.\qquad \left\{\begin{array}{c} \sigma=\rho=0\\ \tau=\frac{\pi}{2},\frac{3\pi}{2}\end{array}\right.\qquad \left\{\begin{array}{c} \tau=\rho=0\\ \sigma=\frac{\pi}{2},\frac{3\pi}{2}\end{array}\right.
\end{equation}
Again we follow the technique of Appendix A and find the set of solutions listed below
\begin{description}
\item{(i)} $\rho=\frac{\pi}{3}$, $\sigma=\tau=0$
\begin{equation}
\begin{split}
& G_1=\naw{\lambda_1-\sqrt{3}\lambda_2+\frac{2}{\sqrt{3}}\lambda_8}\otimes I-I\otimes \naw{\lambda_1-\sqrt{3}\lambda_2+\frac{2}{\sqrt{3}}\lambda_8}\\
&G_2=\naw{\lambda_3+\lambda_4-\sqrt{3}\lambda_5-\frac{1}{\sqrt{3}}\lambda_8}\otimes I-I\otimes \naw{\lambda_3+\lambda_4-\sqrt{3}\lambda_5-\frac{1}{\sqrt{3}}\lambda_8}\\
& G_3=\naw{\lambda_3+2\lambda_6+\frac{1}{\sqrt{3}}\lambda_8}\otimes I-I\otimes\naw{\lambda_3+2\lambda_6+\frac{1}{\sqrt{3}}\lambda_8}\\
& G_4=\naw{\sqrt{3}\lambda_1+\lambda_2-\sqrt{3}\lambda_4-\lambda_5-2\lambda_7}\otimes I+\\& \qquad +I\otimes\naw{\sqrt{3}\lambda_1+\lambda_2-\sqrt{3}\lambda_4-\lambda_5-2\lambda_7}
\end{split}
\end{equation}
\item{(ii)} $\rho=\pi$, $\sigma=\tau=0$
\begin{equation}
\begin{split}
& G_1=\naw{\lambda_1-\frac{1}{\sqrt{3}}\lambda_8}\otimes I-I\otimes\naw{\lambda_1-\frac{1}{\sqrt{3}}\lambda_8}\\
& G_2=\naw{-\lambda_3+2\lambda_4+\frac{1}{\sqrt{3}}\lambda_8}\otimes I-I\otimes\naw{-\lambda_3+2\lambda_4+\frac{1}{\sqrt{3}}\lambda_8}\\
& G_3=\naw{\lambda_3+2\lambda_6+\frac{1}{\sqrt{3}}\lambda_8}\otimes I-I\otimes\naw{\lambda_3+2\lambda_6+\frac{1}{\sqrt{3}}\lambda_8}\\
& G_4=\naw{\lambda_2-\lambda_5+\lambda_7}\otimes I+I\otimes\naw{\lambda_2-\lambda_5+\lambda_7}
\end{split}
\end{equation}
\item{(iii)} $\sigma=\frac{\pi}{2}$, $\rho=\tau=0$
\begin{equation}
\begin{split}
& G_1= \naw{\lambda_1-\lambda_2-\lambda_5-\lambda_7+\frac{1}{\sqrt{3}}\lambda_8}\otimes I-I\otimes\naw{\lambda_1-\lambda_2-\lambda_5-\lambda_7+\frac{1}{\sqrt{3}}\lambda_8}\\
& G_2=\naw{-\lambda_3+2\lambda_4+\frac{1}{\sqrt{3}}\lambda_8}\otimes I-I\otimes\naw{-\lambda_3+2\lambda_4+\frac{1}{\sqrt{3}}\lambda_8}\\
& G_3=\naw{2\lambda_2-\lambda_3+2\lambda_5+2\lambda_6+2\lambda_7-\frac{1}{\sqrt{3}}\lambda_8}\otimes I-\\
&\qquad-I\otimes \naw{2\lambda_2-\lambda_3+2\lambda_5+2\lambda_6+2\lambda_7-\frac{1}{\sqrt{3}}\lambda_8}\\
& G_4=\naw{2\lambda_1+\lambda_2+\lambda_3+\lambda_5-2\lambda_6+\lambda_7+\sqrt{3}\lambda_8}\otimes I+\\
&\qquad +I\otimes\naw{2\lambda_1+\lambda_2+\lambda_3+\lambda_5-2\lambda_6+\lambda_7+\sqrt{3}\lambda_8}
\end{split}
\end{equation}
\item{(iv)} $\tau=\frac{\pi}{2}$, $\rho=\sigma=0$
\begin{equation}
\begin{split}
& G_1=\naw{\lambda_1-\frac{1}{\sqrt{3}}\lambda_8}\otimes I-I\otimes\naw{\lambda_1-\frac{1}{\sqrt{3}}\lambda_8}\\
& G_2=\naw{\lambda_4+\lambda_6-\frac{1}{\sqrt{3}}\lambda_8}\otimes I-I\otimes\naw{\lambda_4+\lambda_6-\frac{1}{\sqrt{3}}\lambda_8}\\
& G_3=\naw{\lambda_2-\lambda_3+\lambda_5+2\lambda_6-\lambda_7-\frac{1}{\sqrt{3}}\lambda_8}\otimes I-\\
& \qquad -I\otimes\naw{\lambda_2-\lambda_3+\lambda_5+2\lambda_6-\lambda_7-\frac{1}{\sqrt{3}}\lambda_8}\\
&G_4=\naw{\lambda_2+\lambda_3+\lambda_4+\lambda_5-\lambda_6-\lambda_7}\otimes I+I\otimes \naw{\lambda_2+\lambda_3+\lambda_4+\lambda_5-\lambda_6-\lambda_7}.
\end{split}
\end{equation}
\end{description}
In all cases there are, as expected, four independent generators.

For generic values of $\rho$, $\sigma$ and $\tau$, which correspond to three different eigenvalues of the reduced density matrix (\ref{d3}), we find two commuting generators. We give few examples:
\begin{description}
\item{(i)} $\rho=\frac{\pi}{2}$, $\sigma=\tau=0$
\begin{equation}
\begin{split}
& G_1=\naw{2\lambda_1-4\lambda_2+\lambda_3+2\lambda_4-4\lambda_5+\frac{1}{\sqrt{3}}\lambda_8}\otimes I-\\
&\qquad -I\otimes\naw{2\lambda_1-4\lambda_2+\lambda_3+2\lambda_4-4\lambda_5+\frac{1}{\sqrt{3}}\lambda_8}\\
& G_2=\naw{\lambda_3+2\lambda_6+\frac{1}{\sqrt{3}}\lambda_8}\otimes I-I\otimes\naw{\lambda_3+2\lambda_6+\frac{1}{\sqrt{3}}\lambda_8}
\end{split}
\end{equation}
\item{(ii)} $\sigma=\pi$, $\rho=\tau=0$
\begin{equation}
\begin{split}
& G_1=\naw{2\lambda_1-3\lambda_3+2\lambda_6+\sqrt{3}\lambda_8}\otimes I-I\otimes\naw{2\lambda_1-3\lambda_3+2\lambda_6+\sqrt{3}\lambda_8}\\
& G_2=\naw{-\lambda_3+2\lambda_4+\frac{1}{\sqrt{3}}\lambda_8}\otimes I-I\otimes\naw{-\lambda_3+2\lambda_4+\frac{1}{\sqrt{3}}\lambda_8}
\end{split}
\end{equation}
\item{(iii)} $\tau=\pi$, $\rho=\sigma=0$
\begin{equation}
\begin{split}
& G_1=\naw{\lambda_1-\frac{1}{\sqrt{3}}\lambda_8}\otimes I-I\otimes\naw{\lambda_1-\frac{1}{\sqrt{3}}\lambda_8}\\
& G_2=\naw{\lambda_4+\lambda_6-\sqrt{3}\lambda_8}\otimes I-I\otimes\naw{\lambda_4+\lambda_6-\sqrt{3}\lambda_8}.
\end{split}
\end{equation}
\end{description}
Again, these results agree with the conclusion that the generic stability subgroup is $S\naw{U(1)\times U(1)\times U(1)}$.

\section{Conclusions}
We have considered the three-strategies ELW game. It is assumed that the set of pure strategies of each player consists of all $SU(3)$ elements. We are looking for the most general gate operator $\hat{J}$ such that the game accomodates all classical pure strategies. The main conclusion is that, with some relabelling of strategies, the gate operator is the exponent of a linear combination of symmetrized tensor products of the generators of Cartan subalgebra of $SU(3)$; it depends on three parameters $\rho$, $\sigma$ and $\tau$. The properties of the game depend on the stability subgroup of the initial state; the stability subgroup is, in turn, determined by the degree of entanglement of the latter. Namely, it depends on the number of equal eigenvalues of the reduced density matrix of initial state. The case of maximal entanglement is particularly interesting; then the stability group is $SU(3)$ group isomorphic to the diagonal subgroup of the group $SU(3)\times SU(3)$ of strategies of $\underline{\text{both}}$ players. As a result, to any strategy of the (say) first player there exists an appropriate counterstrategy of the second one. If the reduced density matrix has two equal eigenvalues, the stability subgroup is $S\naw{U(2)\times U(1)}$; in the generic case of three different eigenvalues it is $S\naw{U(1)\times U(1)\times U(1)}$.

The $SU(2)$-based two-strategies ELW game has additional nice property: all mixed classical strategies can be also represented by pure quantum ones. This is not the case for three strategies game as we show in Appendix C. Such a property is, however, not a crucial one. Once pure classical startegies are property implemented  by pure quantum strategies, the mixed quantum strategies include also the mixed classical ones.

\begin{appendices}
\section{}
We are looking for the stability subgroup of the vector $\widetilde{J}\naw{V^+\otimes V^+}\ket{1,1}$, i.e. for all pairs of matrices $\widetilde{U}_A$, $\widetilde{U}_B$ such that
\eq{\naw{\widetilde{U}_A\otimes \widetilde{U}_B}\widetilde{J}\ket{\widetilde{1},\widetilde{1}}=\widetilde{J}\ket{\widetilde{1},\widetilde{1}}.}
The generators of $\widetilde{U}_A\otimes\widetilde{U}_B$ have the form
\eq{X\otimes I+I\otimes Y}
where $X$ and $Y$ are linear combinations of Gell-Mann matrices. Therefore, we demand
\eq{\naw{X\otimes I+I\otimes Y}\widetilde{J}\ket{\widetilde{1},\widetilde{1}}=0}
or
\eq{\widetilde{J}^{-1}\naw{X\otimes I+I\otimes Y}\widetilde{J}\ket{\widetilde{1},\widetilde{1}}=0.}
Now, noting that
\eq{\varsigma\widetilde{J}\ket{\widetilde{1},\widetilde{1}}=\widetilde{J}\ket{\widetilde{1},\widetilde{1}}}
we conclude that the Lie algebra of stability subgroup is spanned by the eigenvectors of $\varsigma$, i.e. the relevant generators can be chosen in the form 
\eq{X\otimes I\pm I\otimes X.}
Therefore, it is sufficient to solve 
\eq{\widetilde{J}^{-1}\naw{X\otimes I\pm I\otimes X}\widetilde{J}\ket{\widetilde{1},\widetilde{1}}=0.}
In order to compute $\widetilde{J}^{-1}\naw{X\otimes I\pm I\otimes X}\widetilde{J}$ we consider
\eq{ Y\naw{\alpha}\equiv e^{-i\alpha\naw{A\otimes\Lambda}}\naw{X\otimes I}e^{i\alpha\naw{A\otimes\Lambda}}\label{c}}
\eq{Z\naw{\alpha}=e^{-i\alpha\naw{A\otimes\Delta}}\naw{X\otimes I}e^{i\alpha\naw{A\otimes\Delta}}\label{cc}}
where $A$ is an element of Cartan subalgebra of $SU(3)$. With an appropriate choice of the basis we have 
\eq{\com{A,X}=a\naw{X}X.\label{c3}}
The matrices $\Lambda$ and $\Delta$ obey
\eq{\Lambda^3-\Lambda=0,\qquad \Delta^3-\Delta=0.}
Using this and the Hausdorff formula one finds
\eq{Y\naw{\alpha}=Y_1\naw{\alpha}\otimes I+Y_2\naw{\alpha}\otimes \Lambda+Y_3\naw{\alpha}\otimes\Lambda^2.\label{c1}}
Eq. (\ref{c}) implies
\eq{\dot{Y}\naw{\alpha}=-i\naw{\com{A,Y_1}\otimes\Lambda+\com{A,Y_2}\otimes\Lambda^2+\com{A,Y_3}\otimes\Lambda}\label{c2}}
or, comparing eqs. (\ref{c1}) and (\ref{c2})
\eq{\begin{split}
& \dot{Y}_1\naw{\alpha}=0\\
& \dot{Y}_2\naw{\alpha}=-i\naw{\com{A,Y_1}+\com{A,Y_3}}\\
& \dot{Y}_3\naw{\alpha}=-i\com{A,Y_2}
\end{split}.}
So we get
\eq{\begin{split}
&Y_1\naw{\alpha}=X\\
& Y_2\naw{\alpha}=\frac{1}{2}\naw{e^{-i\alpha A}X e^{i\alpha A}-e^{i\alpha A}X e^{-i\alpha A}}\\
& Y_3\naw{\alpha}=\frac{1}{2}\naw{e^{-i\alpha A}X e^{i\alpha A}+e^{i\alpha A}X e^{-i\alpha A}-2X}
\end{split}.}
By virtue of eq. (\ref{c3}) we find finally
\eq{\begin{split}
& Y_1\naw{\alpha}=X\\
& Y_2\naw{\alpha}=-i\sin\naw{\alpha a\naw{X}}X\\
& Y_3\naw{\alpha}=\naw{\cos\naw{\alpha a\naw{X}}-1}X
\end{split}} 
and
\eq{Y\naw{\alpha}=X\otimes\naw{I-i\sin\naw{\alpha a\naw{X}}\Lambda+\naw{\cos\naw{\alpha a\naw{X}}-1}\Lambda^2}.\label{f7}}
Similarly
\eq{Z\naw{\alpha}=X\otimes\naw{I-i\sin\naw{\alpha a\naw{X}}\Delta+\naw{\cos\naw{\alpha a\naw{X}}-1}\Delta^2}.\label{f8}}
Let us put
\eq{\begin{split}
& \tau\Lambda\otimes\Lambda+\rho\naw{\Lambda\otimes\Delta+\Delta\otimes\Lambda}+\sigma\Delta\otimes\Delta=\\
& =\naw{\tau\Lambda+\rho\Delta}\otimes\Lambda+\naw{\rho\Lambda+\sigma\Delta}\otimes\Delta \equiv A_1\otimes\Lambda+A_2\otimes\Delta.\end{split}}
Therefore
\eq{\widetilde{J}=e^{iA_1\otimes\Lambda}e^{iA_2\otimes\Delta};}
using eqs (\ref{f7}) and (\ref{f8}) we find
\eq{\begin{split}
& \widetilde{J}^{-1}\naw{X\otimes I}\widetilde{J}=e^{-iA_2\otimes\Delta}e^{-iA_1\otimes\Lambda}\naw{X\otimes I}e^{iA_1\otimes\Lambda}e^{iA_2\otimes\Delta}=\\& =e^{-iA_2\otimes\Delta}\naw{X\otimes\naw{I-i\sin\naw{a_1\naw{X}}\Lambda+\naw{\cos\naw{a_1\naw{X}}-1}\Lambda^2}}e^{iA_2\otimes\Delta}=\\
& =e^{-iA_2\otimes\Delta}\naw{X\otimes I}e^{iA_2\otimes\Delta}\naw{I\otimes\naw{I-i\sin\naw{a_1\naw{X}}\Lambda+\naw{\cos\naw{a_1\naw{X}}-1}\Lambda^2}}=\\
& =X\otimes\naw{I-i\sin\naw{a_2\naw{X}}\Delta+\naw{\cos\naw{a_2\naw{X}}-1}\Delta^2}\big( I-i\sin\naw{a_1\naw{X}}\Lambda+\\
&+\naw{\cos\naw{a_1\naw{X}}-1}\Lambda^2\big)=\\
&=X\otimes\naw{\naw{I-is_2\Delta+\naw{c_2-1}\Delta^2}\naw{I-is_1\Lambda+\naw{c_1-1}\Lambda^2}}
\end{split}}
where $s_i\equiv\sin\naw{a_i\naw{X}}$, $c_i\equiv\cos\naw{a_i\naw{X}}$. Summarizing, the following relation should hold for the generators of stability subgroup
\eq{\naw{X\otimes\Omega\pm\Omega\otimes X}\naw{\ket{\widetilde{1}}\otimes\ket{\widetilde{1}}}=0}
where $\Omega$ is the matrix of the form
\eq{\Omega=\left(\begin{array}{ccc}
e^{-i\naw{a_1+a_2}} & 0 & 0\\
0 & e^{ia_1} & 0\\
0 & 0 & e^{ia_2} \end{array}\right).}

\section{} 
Let us determine the values of the parameter $\tau$, $\rho$, $\sigma$ corresponding to maximal entanglement. The reduced density matrix $\text{Tr}_B\rho_i$ reads
\begin{equation}
\text{Tr}_B\rho_i =\frac{1}{9}\left(\begin{array}{c|c|c}
 &  e^{i\naw{3\rho+\sigma+2\tau}}+ & e^{i\naw{3\rho+2\sigma+\tau}}+\\
 3 & +e^{-i\naw{\rho +2\tau}}+ & +e^{-i\naw{2\rho+\tau}}\\
 & +e^{-i\naw{2\rho+\sigma}} & +e^{-i\naw{\rho+2\sigma}}\\ \hline
  e^{-i\naw{3\rho+\sigma+\tau}}+ &  & e^{i\naw{\sigma-\tau}}+\\
  +e^{i\naw{\rho+2\tau}}+ & 3 & +e^{-i\naw{\rho-\tau}}+\\
  +e^{2\rho+\sigma} & & +e^{i\naw{\rho-\sigma}}\\ \hline
  e^{-i\naw{3\rho+2\sigma+\tau}}+ & e^{-i\naw{\sigma-\tau}} + & \\
  + e^{i\naw{2\rho+\tau}}+ & +e^{i\naw{\rho-\tau}}+ & 3\\
  +e^{i\naw{\rho+2\sigma}} & +e^{-i\naw{\rho-\sigma}} & \\ 
  \end{array}\right)\label{dd2} 
 \end{equation}
 The vainishing of off-diagonal components yield
 \begin{equation}
 e^{i\naw{\alpha+\beta}}+e^{-i\alpha}+e^{-i\beta}=0\label{dd1}
 \end{equation}
 for $\alpha=\rho+2\tau$, $\beta=\sigma+2\rho$, $\alpha=2\rho+\tau$, $\beta=2\sigma+\rho$ and $\alpha=\tau-\rho$, $\beta=\rho-\sigma$. \\
 Eq. (\ref{dd1}) gives $\modu{e^{i\alpha}+e^{i\beta}}=1$ or 
 \begin{equation}
 \cos\naw{\alpha-\beta}=-\frac{1}{2}\quad i.e. \quad \alpha-\beta=\pm\frac{2\pi}{3}+2k\pi.
 \end{equation}
Inserting this back into eq. (\ref{dd1}) one arrives at six solutions (modulo $2k\pi$):
\begin{equation}\begin{split}
& (i)\quad \alpha=0,\quad \beta=\pm\frac{2\pi}{3}\\
&(ii)\quad\alpha=\pm\frac{2\pi}{3},\quad\beta=0\\
&(iii)\quad\alpha=\pm\frac{2\pi}{3}\quad\beta=\mp\frac{2\pi}{3}.
\end{split}\label{dd3}
\end{equation}
Considering the $(2,3)$-element of the matrix (\ref{dd2}) we have 
\begin{equation}
\begin{split}
& \alpha=\tau-\rho\\
& \beta=\rho-\sigma.
\end{split}
\end{equation}
Inserting here for $\alpha$ and $\beta$ the solutions (\ref{dd3}) we find $\rho$ and $\tau$ in terms of $\sigma$. This allows to determine $\sigma$ from the condition that one of the remaining off-diagonal element vanishes; it remains to check that the third element also vanishes. In this way we obtain the following solutions:
\begin{equation}\begin{split}
& \left\{ \begin{array}{c}
\tau=\rho=\sigma-\frac{2\pi}{3}\\
\sigma=\frac{2\pi}{3},\frac{8\pi}{9},\frac{10\pi}{9},\frac{4\pi}{3},\frac{14\pi}{9},\frac{16\pi}{9},2\pi \end{array}\right.\\
& \left\lbrace \begin{array}{c}
\tau=\rho=\sigma+\frac{2\pi}{3}\\
\sigma=0,\frac{2\pi}{9},\frac{4\pi}{9},\frac{2\pi}{3},\frac{8\pi}{9},\frac{10\pi}{9},\frac{4\pi}{3},\frac{14\pi}{9},\frac{16\pi}{9} \end{array}\right.\\
& \left\{ \begin{array}{c}
\tau=\sigma-\frac{2\pi}{3}\\
\rho=\sigma=\frac{2\pi}{3},\frac{8\pi}{9},\frac{10\pi}{9},\frac{4\pi}{3},\frac{14\pi}{9},\frac{16\pi}{9},2\pi\end{array}\right. \\
&\left\lbrace \begin{array}{c}
\tau=\sigma+\frac{2\pi}{3}\\
\rho=\sigma=0,\frac{2\pi}{9},\frac{4\pi}{9},\frac{2\pi}{3},\frac{8\pi}{9},\frac{10\pi}{9},\frac{4\pi}{3},\frac{14\pi}{9},\frac{16\pi}{9} \end{array}\right.\\
& \left\{ \begin{array}{c}
\rho=\sigma-\frac{2\pi}{3}\\
\tau=\sigma=\frac{2\pi}{3},\frac{8\pi}{9},\frac{10\pi}{9},\frac{4\pi}{3},\frac{14\pi}{9},\frac{16\pi}{9},2\pi \end{array}\right.\\
& \left\lbrace \begin{array}{c}
\rho=\sigma+\frac{2\pi}{3}\\
\tau=\sigma=0,\frac{2\pi}{9},\frac{4\pi}{9},\frac{2\pi}{3},\frac{8\pi}{9},\frac{10\pi}{9},\frac{4\pi}{3},\frac{14\pi}{9},\frac{16\pi}{9} \end{array}\right.
\end{split}
\end{equation}
Consider next the case of partial entanglement, i.e. the case when the matrix (\ref{dd2}) has two equal eigenvalues. In order to find the constraint on $\rho$, $\sigma$ and $\tau$ one can neglect the diagonal part of (\ref{dd2}) and consider the characteristic equation
\begin{equation}
\det\left(\begin{array}{ccc}
-\lambda & a & b\\
\overline{a} & -\lambda & c\\
\overline{b} & \overline{c} & -\lambda \end{array}\right)=0\label{dd4}
\end{equation}
where $a$, $b$ and $c$ are the off-diagonal elements of (\ref{dd2}) ($a=e^{i\naw{3\rho+\sigma+2\tau}}+e^{-i\naw{\rho+2\tau}}+e^{-i\naw{2\rho+\sigma}}$, etc.). 
Eq. (\ref{dd4}) yields
\begin{equation}
\lambda^3-\naw{\modu{a}^2+\modu{b}^2+\modu{c}^2}\lambda-\naw{a\overline{b}c+\overline{a}b\overline{c}}=0.\label{dd5}
\end{equation}
If (\ref{dd5}) has a double root
\begin{equation}
3\lambda^2-\naw{\modu{a}^2+\modu{b}^2+\modu{c}^2}=0\label{dd6}
\end{equation}
or
\begin{equation}
\lambda=\pm\sqrt{\frac{\modu{a}^2+\modu{b}^2+\modu{c}^2}{3}}.\label{dd7}
\end{equation}
Inserting this back into (\ref{dd5}) one obtains
\begin{equation}
\mp\frac{2}{3}\naw{\modu{a}^2+\modu{b}^2+\modu{c}^2}\sqrt{\frac{\modu{a}^2+\modu{b}^2+\modu{c}^2}{3}}=a\overline{b}c+\overline{a}b\overline{c}
\end{equation}
which hold for at least one choice of sign on the left hand side. Taking a square of both sides yields 
\begin{equation}
\frac{4}{27}\naw{\modu{a}^2+\modu{b}^2+\modu{c}^2}^3=4\text{Re}\naw{a\overline{b}c}^2.\label{f9}
\end{equation}
Due to the inequality
\begin{equation}
\frac{1}{3}\naw{\modu{a}^2+\modu{b}^2+\modu{c}^2}\geq\sqrt[3]{\modu{a}^2\modu{b}^2\modu{c}^2}
\end{equation}
which is saturated iff $\modu{a}=\modu{b}=\modu{c}$, one finds
\begin{equation}
\frac{1}{27}\naw{\modu{a}^2+\modu{b}^2+\modu{c}^2}^3\geq\modu{a}^2\modu{b}^2\modu{c}^2\geq\modu{a}^2\modu{b}^2\modu{c}^2\cos^2\alpha
\end{equation}
where $\alpha=\arg a-\arg b+\arg c$. Therefore, eq. (\ref{f9}) holds only if $\modu{a}^2=\modu{b}^2=\modu{c}^2$, $\arg a-\arg b+\arg c=0,\pi\naw{mod 2\pi}$. Then, denoting by $\lambda_0$ a duble root, one finds
\begin{equation}
\modu{a}^2=\modu{b}^2=\modu{c}^2=\lambda_0.
\end{equation}
The third root equals $-2\lambda_0$.

Due to the complicated structure of the elements $a$, $b$, $c$, when expressed in terms of basic parameters $\rho$, $\sigma$, $\tau$, we solve eqs. (\ref{dd6}), (\ref{dd7}) in the special case of only one nonvanishing parameter. The resulting solutions read:
\begin{equation}
 \left\{\begin{array}{c}
\sigma=\tau=0\\
\rho=\frac{\pi}{3},\pi,\frac{5\pi}{3}
\end{array}\right.\qquad \left\{\begin{array}{c} \sigma=\rho=0\\ \tau=\frac{\pi}{2},\frac{3\pi}{2}\end{array}\right.\qquad \left\{\begin{array}{c} \tau=\rho=0\\ \sigma=\frac{\pi}{2},\frac{3\pi}{2}\end{array}\right. .
\end{equation}

\section{}
We solve here the problem whether all classical mixed strategies can be implemented by pure quantum ones. In order to preserve the factorization property for probabilities the strategy of any player must be of the form
\begin{equation}
U=e^{i\naw{\alpha\Lambda+\beta\Delta}}
\end{equation}
The relevant probabilities of respective strategies read
\begin{equation}
p_\sigma=\modu{\bra{\widetilde{\sigma}}U\ket{\widetilde{1}}}^2
\end{equation}
or, explicitly,
\begin{equation}
\begin{split}
& p_1=\frac{1}{9}\modu{e^{i\alpha}+e^{-i\alpha+i\beta}+e^{-i\beta}}^2\\
& p_2=\frac{1}{9}\modu{e^{i\alpha}+\varepsilon^2e^{-i\alpha+i\beta}+\varepsilon e^{-i\beta}}^2\\
&  p_3=\frac{1}{9}\modu{\varepsilon^2e^{i\alpha}+e^{-i\alpha+i\beta}+\varepsilon e^{-i\beta}}^2.
\end{split}
\end{equation}
Let us call $e^{-i\naw{\alpha+\beta}}\equiv u_1$, $e^{i\naw{\beta-2\alpha}}\equiv u_2$, then
\begin{equation}
\begin{split}
& p_1=\frac{1}{9}\modu{1+u_1+u_2}^2\\
& p_2=\frac{1}{9}\modu{1+\varepsilon u_1+\varepsilon^2u_2}^2.
\end{split}\label{e}
\end{equation}
Now, $p_{1,2}$ obey $0\leq p_{1,2}\leq 1$, $0\leq p_1+p_2\leq 1$.

Let $\gamma=\frac{1}{2}\naw{\arg u_1-\arg u_2}$ (if $\gamma>\frac{\pi}{2}$ we take $\gamma\rightarrow\pi-\gamma$) and $\delta=\arg\naw{u_1+u_2}$. Then eqs. (\ref{e}) can be rewritten as
\begin{equation}
\begin{split}
& \cos^2\gamma+\cos\gamma\cos\delta=\lambda\equiv\frac{9p_1-1}{4}\\
& \cos^2\naw{\gamma+\frac{2\pi}{3}}+\cos\naw{\gamma+\frac{2\pi}{3}}\cos\delta=\mu\equiv\frac{9p_2-1}{4}
\end{split}
\end{equation}
and $-\frac{1}{4}\leq\lambda,\mu\leq 2$, $-\frac{1}{2}\leq\lambda+\mu\leq\frac{7}{4}$.
Eliminating $\cos\delta$ through
\begin{equation}
\cos\delta=\frac{\lambda-\cos^2\gamma}{\cos\gamma}\label{e1}
\end{equation}
we find cubic equation for $\text{tg}\gamma$
\begin{equation}
\naw{3-2\lambda-\mu}+2\sqrt{3}\naw{2-\lambda}\text{tg}\gamma+\naw{3-2\lambda-\mu}\text{tg}^2\gamma-2\lambda\sqrt{3}\text{tg}^3\gamma=0
\end{equation}
Solving the last equation we find $\gamma$ and then $\cos\delta$ from eq. (\ref{e1}). The solution exists if $-1\leq\cos\delta\leq 1$. One can check numerically that, in general, this is not the case. For example, taking $\lambda=-\frac{1}{8}$ and $\mu=1$ we obtain that the right hand side of eq. (\ref{e1}) is equal to $-1,12041$. 
\end{appendices}

\subsection*{Acknowledgement}
I would like to thank Professor Piotr Kosi\'nski (Department of Computer Science, Faculty of Physics and Applied Informatics, University of L\'od\'z, Poland) for helpful discussion and useful remarks. 
 This research is supported by the NCN Grant no. DEC-2012/05/D/ST2/00754.


\begin{thebibliography}{99}
\bibitem{EisertWL} J. Eisert, M. Wilkens, M. Lewenstein, \emph{Phys. Rev. Lett.} \textbf{83} (1999), 3077
\bibitem{EisertW} J. Eisert, M. Wilkens, \emph{J. Mod. Opt.} \textbf{47} (2000), 2543
\bibitem{Meyer} D. Meyer, \emph{Phys. Rev. Lett.} \textbf{82} (1999), 1052
\bibitem{Marinatto} L. Marinatto, T. Weber, \emph{Phys. Lett}, \textbf{A272} (2000), 291
\bibitem{Benjamin} S. Benjamin, \emph{Phys. Lett.}, \textbf{A277} (2000), 180
\bibitem{MarinattoWeb} L. Marinatto, T. Weber, \emph{Phys. Lett.} \textbf{A 277} (2000), 183
\bibitem{BenjaminHay} S. Benjamin, P. Hayden, \emph{Phys. Rev. Lett.} \textbf{87(6)} (2001), 069801
\bibitem{Iqbal} A. Iqbal, A. Toor, \emph{Phys. Lett} \textbf{A280} (2001), 249
\bibitem{DuLi} J. Du, H. Li,X. Xu, X. Zhou, R. Han, \emph{Phys. Lett} \textbf{A289} (2001), 9
\bibitem{EisertWL1} J. Eisert, M. Wilkens, M. Lewenstein, \emph{Phys. Rev.Lett.} \textbf{87} (2001), 069802
\bibitem{FlitneyA} A. Flitney, D. Abbott, \emph{Fluct. Noise Lett.} \textbf{2} (2000), R175
\bibitem{Iqbal1} A. Iqbal, A. Toor, \emph{Phys. Rev.} \textbf{A65} (2002), 052328
\bibitem{DuLi1} J. Du, H. Li, X. Xu, M. Shi, J. Wu, X. Zhou, R. Han, \emph{Phys. Rev. Lett.} \textbf{88} (2002), 137902
\bibitem{Enk} S.J. van Enk, R. Pike, \emph{Phys. Rev} \textbf{A66} (2002), 024306
\bibitem{DuLi2} J. Du, X. Xu, H. Li, X. Zhou, R. Han, \emph{Fluct. Noise Lett.} \textbf{2} (2002), R189
\bibitem{FlitneyA1} A. Flitney, D. Abbott, \emph{Proc. R. Soc. Lond.} \textbf{A459} (2003), 2463
\bibitem{PiotrowskiS} E. Piotrowski, J. Sladkowski, \emph{Int. Journ. Theor. Phys.} \textbf{42} (2003), 1089
\bibitem{DuLi3} J. Du, H. Li, X. Xu, X. Zhou, R. Han, \emph{Journ. Phys.} \textbf{A36} (2003), 6551
\bibitem{Zhou} L. Zhou, L. Kuang, \emph{Phys. Lett} \textbf{A315} (2003), 426
\bibitem{Chen} L. Chen, H. Ang, D. Kiang, L. Kwek, C. Lo, \emph{Phys. Lett.} \textbf{A316} (2003), 317
\bibitem{Lee} C.F. Lee, N.F. Johnson, \emph{Phys. Rev.} \textbf{A67} (2003), 022311
\bibitem{Shimamura} J. Shimamura, S. Ozdemir, F. Morikoshi, N. Imoto, \emph{Int. Journ. Quant. Inf.} \textbf{2} (2004), 79
\bibitem{Landsburg} S. Landsburg, \emph{Notices of the Am. Math. Soc.} \textbf{51} (2004), 394
\bibitem{Rosero} A. Rosero, "Classification of Quantum Symmetric Non-zero Sum $2\times 2$ Games in the Eisert Scheme", quant-phys/0402117
\bibitem{NawazT1} A. Nawaz, A. Toor, \emph{Journ. Phys.} \textbf{A37} (2004), 11457
\bibitem{NawazT2} A. Nawaz, A. Toor, \emph{Journ. Phys.} \textbf{A37} (2004), 4437
\bibitem{Iqbal2} A. Iqbal, "Studies in the theory of quantum games", quant-phys/0503176
\bibitem{FlitneyA2} A. Flitney, D. Abbott, \emph{Journ. Phys.} \textbf{A38} (2005), 449
\bibitem{Ichikawa} T. Ichikawa, I. Tsutsui, \emph{Ann. Phys.} \textbf{322} (2007), 531
\bibitem{Cheon} T. Cheon, I. Tsutsui, \emph{Phys. Lett.} \textbf{A348} (2006), 147
\bibitem{Patel} N. Patel, \emph{Nature} \textbf{445} (2007), 144
\bibitem{Ichikawa1} T. Ichikawa, I. Tsutsui, \emph{Journ. Phys. A: Math. and Theor.} \textbf{41} (2008), 135303
\bibitem{FlitneyA3} A. Flitney, L. Hollenberg, \emph{Phys. Lett.} \textbf{A363} (2007), 381
\bibitem{Nawaz3} A. Nawaz, "The generalized quantization schemes for games and its application to quantum information", arXiv:1012.1933
\bibitem{Landsburg1} S. Landsburg, \emph{Proc. Am. Math. Soc.} \textbf{139} (2011), 4413
\bibitem{Landsburg2} S. Landsburg, \emph{Wiley Encyclopedia of operations Research and Management science} (2011)
\bibitem{Schneider2} D. Schneider, \emph{Journ. Phys.} \textbf{A44} (2011), 095301
\bibitem{Schneider} D. Schneider, \emph{Journ. Phys.} \textbf{A45} (2012), 085303
\bibitem{Avishai} Y. Avishai, "Some Topics in Quantum Games", arXiv:1306.0284
\bibitem{Bolonek} K. Bolonek-Laso\'n, P. Kosi\'nski, \emph{Prog. Theor. Exp. Phys.} (2013), 073A02
\bibitem{Ramzan1} M. Ramzan, \emph{ Quant. Inf. Process.} \textbf{12} (2013), 577 \bibitem{Ramzan2} M. Ramzan, M. K. Khan, \emph{Fluctuation and Noise Letters} \textbf{12} (2013), 1350025 
\bibitem{Nawaz4} A. Nawaz, \emph{Chin. Phys. Lett.} \textbf{30(5)} (2013), 050302
\bibitem{Frackiewicz} P. Frackiewicz, \emph{Acta Phys. Polonica B} \textbf{44} (2013), 29
\bibitem{Nawaz5} A. Nawaz, "Werner-like States and Strategies form of Quantum games", arXiv:1307.5508
\bibitem{Nawaz6} A. Nawaz, \emph{J. Phys. A: Math. Theor. J. Phys.} \textbf{45} (2012), 195304
\bibitem{Avishay} Y. Avishai, arXiv:1402.1982 (quant.-ph.)
\bibitem{Bolonek1} K. Bolonek-Laso\'n, in preparation
\bibitem{Bolonek2} K. Bolonek-Laso\'n, arXiv:1402.3932
\bibitem{Bolonek3} K. Bolonek-Laso\'n, in preparation

\end{thebibliography}
\end{document}